\documentstyle[aps,prd,preprint,tighten]{revtex}
\begin{document}
\draft
\title{Sterile neutrinos as a solution to all neutrino anomalies
\thanks{hep-ph/9506228}
}
\author{Juha T. Peltoniemi \thanks{peltonie@sissa.it} }
\address{ Istituto Nazionale di Fisica Nucleare, Sezione di Trieste,\\
and International School for Advanced
Studies, 34013 Trieste, Italy}
\date{\today}
\maketitle
\begin{abstract}
The sterile neutrino solutions to different irregular results and observations
in neutrino physics are studied. It is pointed out that
introducing sterile neutrinos helps to solve simultaneously the observed
anomalies. It is argued that sterile neutrinos can solve
the conflict between dark matter neutrinos, LSND result and supernova
nucleosynthesis. Other supernova constraints
for sterile neutrinos are revised.
Possibilities to avoid the big bang nucleosynthesis constraints for sterile
neutrinos are explored. It is claimed that sterile neutrinos can solve the
crisis in big bang nucleosynthesis. It is pointed out that sterile neutrinos
can provide a consistent explanation to the anomalies observed at Karmen.
It is argued that sterile neutrinos are valid dark matter candidates.
It is claimed that the conversions to sterile states are consistent solutions
to both the solar and the atmospheric neutrino problems, and cannot be
ruled out by cosmological or astrophysical arguments.
Models for the masses and the interactions of sterile neutrinos are reviewed.
\end{abstract}

\section{Introduction}

Sterile neutrinos or neutrino-like objects are predicted in many grand
unified theories. Supersymmetric models always involve many new neutral
fermions. Although many of the new neutral fermions are not generally called
neutrinos, being associated with some other particles, in many occasions
they behave as sterile neutrinos, and may even mix with the known
neutrinos. Here   generically any  fermion
without standard model interactions mixing with the ordinary neutrinos
is called a sterile neutrino.

I focus on sterile neutrinos that are light, i.e. have masses in
the same scale as is allowed and expected for the known neutrinos
($m_{\nu_S} < 100$ MeV). There are several plausible ways to generate
appropriate mass matrices, basically the small sterile neutrino mass could
be due to similar mechanisms as that of the ordinary neutrinos, like
see-saw or radiative mechanisms.

The purpose of this work is to find out whether the present astronomical
and terrestrial observations allow, or even require, the existence of light
sterile neutrinos. I will explore the sterile neutrino solutions to
several irregular results, especially I will investigate whether the
introduction of  sterile neutrinos with appropriate properties helps
to build consistent scenarios solving
simultaneously all or most of the observed
anomalies.

Astrophysical constraints for the properties
of sterile neutrinos are reviewed, and their validity is discussed. I present
some new
limits for neutrino interactions involving sterile neutrinos, and revise
some old.
The ways to avoid the constraints are explored,
to gain more parameter space for certain solutions.

It is  argued that sterile neutrino solutions to both the atmospheric neutrino
problem and the solar neutrino problem are consistent with other phenomena,
for all plausible parameter ranges. Especially it is emphasized that they
cannot be ruled out by cosmological arguments.

The anomalous results
occupying the neutrino physicists are recalled in section \ref{anom},
and the sterile neutrino solutions to each of them are discussed,
as well as the possibilities to solve simultaneously
several anomalies, using the sterile neutrino degrees of
freedom. The possible types of desired mass hierarchies
are presented.
It is also pointed out that introducing
sterile neutrinos helps to build consistent scenarios accommodating
the observations of LSND \cite{LSND95}.

In section
\ref{models} several models appropriate for
generating light
sterile neutrinos are reviewed. Some specific types
of mass matrices are investigated more thoroughly.
Section \ref{pheno} is devoted to general phenomenological
aspects of models involving sterile neutrinos. I list
laboratory constraints for sterile neutrinos, and talk about
different decay modes of a massive neutrino. Especially
it is pointed out that the absence of the GIM mechanism can lead
to a substantial decay rate by standard neutral currents.

In section \ref{sn} the role of sterile
neutrinos in supernovae is considered. I show a way to avoid
the limits for the $\nu_\mu \to \nu_e$ oscillations
derived by requiring the supernova to be the
origin of heavy elements \cite{Qianetal93,QianFuller94}. The limits for
the production of sterile neutrinos in the supernova
core, by oscillations or by explicit interactions between
sterile neutrinos and ordinary matter, are discussed.
The limits for the neutrino decay involving sterile states are explored.

The section
\ref{bbn} concentrates on the effects of the sterile neutrinos
in the early universe. The possibilities to avoid the
nucleosynthesis constraints
\cite{BarbieriDolgov90,Kainulainen90a,EnqvistKainulainenThomson92,%
ShiSchrammFields93,EnqvistMaalampiSemikoz95}
for the excitation of sterile states by
oscillation or explicit interactions are discussed.
However, at present the standard particle physics model gives such
a poor fit to the nucleosynthesis model and the observations of the light
elements \cite{Hataetal95} that it is questionable to use cosmological
arguments
to rule out any
non-standard scenario.
It is pointed out that decaying sterile neutrinos could save the
nucleosynthesis.
The role of sterile neutrinos as the dark matter of the universe is discussed
at the end of this section.

Section \ref{karmen} treats the anomalies observed in Karmen
\cite{Armbrusteretal95}.
It is shown that the sterile neutrino solution is not inconsistent with
the supernova dynamics or the big bang nucleosynthesis. A new range
of mixings is suggested, in deviation from \cite{BargerPhillipsSarkar95}.

Finally, in section \ref{conc}
the results are summarized, and some other
implications are discussed.

\section{Sterile neutrinos and irregular results of neutrino physics}
\label{anom}
\subsection{Anomalous observations}

There has been, there is, and there will be several anomalies
related to the neutrino physics. In the following I list some
of them, and discuss briefly the possibility to solve them
individually by sterile neutrinos.

\begin{enumerate}
\item
{\em Solar neutrino problem.} All the four on-going solar neutrino
experiments
\cite{Homestake94,Kamiokande94,Sage94,Gallex94}
have measured a deviation from the theoretical neutrino flux.
Beside the canonical solutions it has been suggested that the deficit
is due to the conversion of the emitted electron neutrinos to
a sterile state, either by vacuum oscillation, resonant
neutrino conversion \cite{Bargeretal91}, or spin flip
\cite{VoloshinVysotskyOkun86a}. The vacuum oscillation solution is not
consistent with the recent data \cite{KrastevPetcov93} while the resonant
conversion solution can
be fitted reasonably well
\cite{KrastevSmirnov94,HataLangacker94} (See also
\cite{KrastevPetcov93,BilenkyGiunti94,BilenkyGiunti95}).
The typical solutions require squared
mass
difference $(4\times 10^{-6} \ldots 10^{-5}) \mbox{ eV}^2$, and mixing $\sin^2
2\theta \sim
4\times 10^{-4} \ldots 4\times 10^{-3}$. The spin flip solution also fits
well \cite{AkhmedovLanzaPetcov95}, but it would require
magnetic
fields higher than usually believed to exist in the sun.
Ignoring the results of
one experiment increases the parameter range.

\item {\em Atmospheric neutrino problem.} The deficit of muon neutrinos
produced in atmosphere has also been observed by several experiments
\cite{Fukudaetal94,Becker-Szendyetal92,Agliettaetal89,Soudan94,%
Daumetal95}.
The oscillation to sterile
neutrinos has been generally considered as an inconsistent solution to
this anomaly, because of
the apparent conflict with the neutrino
mixing bounds from the physics of the early universe
\cite{EnqvistKainulainenThomson92b}.
Ignoring those bounds, the sterile neutrino solution would give
as good a fit as the oscillation to tau neutrinos
\cite{AkhmedovLipariLusignoli92}. The masses considered
are in the range $10^{-3} \mbox{ eV}^2 < \delta m^2 < 0.5 \mbox{ eV}^2$,
while the mixing should be quite large, $\sin^2 \theta > 0.5 $.
The most favored region is close to maximal mixing with
 $10^{-3} \mbox{ eV}^2 < \delta m^2 < 3\times 10^{-3} \mbox{ eV}^2$.

\item {\em Los Alamos neutrinos.}
The Liquid Scintillator Neutrino Detector (LSND) experiment has observed an
excess of electron neutrinos, in
a beam originally
consisting  of muon neutrinos \cite{LSND95}.
The claimed signal is consistent with the oscillation of muon neutrinos
to electron neutrinos, but the results are still too controversy to be
conclusive. If it were a signal of neutrino mixing, it would imply
squared mass differences from $0.2 \mbox{ eV}^2$ to $10 \mbox{ eV}^2$
and above, and a mixing $10^{-3} < \sin^2 2\theta < 0.04$, depending
on the mass scale. For a wide range of parameters the results are consistent
with other experiments (Karmen,
BNL E776, Bugey) which, however, exclude a large part of the parameter space.
Contrary to some previous reports of positive results
for a neutrino mass from other experiments, the speculated mass scale is very
natural, and even
expected. Especially, the results allow neutrino masses in a cosmologically
important scale. Hence
it is well motivated to study the consequences of such
masses, whether or not the preliminary results will be confirmed.

The Los Alamos results cannot be
interpreted as a signal of sterile neutrinos. However, taken that the results
are incompatible (or marginally compatible) with the
solar neutrino
and atmospheric neutrino solutions, a simultaneous solution to all three
cases would require a fourth, necessarily sterile neutrino.

\item {\em Supernova neutrinos.} The supernova SN1987A did not
show any anomalies with neutrinos, instead it gave us a possibility
to gather information on the properties of neutrinos, especially
giving constraints for the sterile neutrinos themselves.
These constraints may restrict the sterile neutrino solutions
to some anomalies.

The problem of the supernovae is that it is difficult to make
them blow up: most simulations end up with a stalled explosion
clearly contrary to all observations. The present status of modelling
favors the scenarios where the neutrino radiation helps to blow up
the envelope. It has been suggested that successive conversions to
a sterile state and then back to active state would help to bring
sufficiently energy to the outer regions to throw them away
\cite{Dar87,Peltoniemi92}.

\item {\em Dark matter.}
Neutrinos have been always considered as the main candidates
for the invisible matter in the universe, required by several
observations. The most popular solution to the formation of
the structures in the universe employs a mixed dark matter model,
where neutrinos with a mass of a few electronvolts provide some
30 \% of the energy density
of the universe, the rest consisting mainly of heavier particles.
Some simulations give the best fit with two neutrinos with a mass
of 2.4 eV \cite{Primacketal95}.

It has been speculated that some part of the dark matter would
consist of sterile neutrinos \cite{Peltoniemi93,DodelsonWidrow94}. It is noted
that
sterile neutrinos
may behave differently than active neutrinos with the same mass,
since their average energies may be different.

As an alternative to the mixed dark matter it has been suggested that
one can do with only cold dark matter, if the onset of the structure
formation is delayed, as would be caused by a massive particle
decaying lately
\cite{BardeenBondEfstathiou87,BondEfstathiou91,%
DodelsonGyukTurner94,WhiteGelminiSilk95}.
Here it is recalled that sterile neutrinos would either
provide a candidate for that lately decaying particle, or provide
a decay channel for an active neutrino.

\item {\em Anomalous ionization of interstellar hydrogen.}
It has been suggested
\cite{deRujulaGlashow80,RephaeliSzalay81,Sciama82,%
MelottMcKayRalston88,Sciama90}
that a neutrino
($\nu_\mu$ or $\nu_\tau$) with
a mass of 27 eV would explain the anomalously large ionization
of interstellar hydrogen by decaying to a lighter neutrino and
a photon. The scenario is in agreement with the
observed constraints for the ultraviolet radiation
\cite{RoosBowyerLamptonPeltoniemi95,BowyerLamptonPeltoniemiRoos95},
but it may not
fit the current understanding of the structure formation of the universe,
without additional assumptions on the initial conditions.

The alternative of using sterile neutrinos as the decaying
particles \cite{Peltoniemi93} fits the astronomical observations
as well, and is also consistent with the big bang nucleosynthesis.
A sterile neutrino of 30 eV is, however,  even worse as  hot
dark matter, but since its abundance may be lower
it allows other coexisting particles to form the main part of
the dark matter. Alternatively a non thermal production of the
sterile states may improve their properties. In \cite{Peltoniemi93}
it was shown that models satisfying all the requirements can be built,
in case the photons are emitted from the decay of one sterile state
to another. A more economic alternative
would be the decay of a sterile state to
an active tau neutrino state.
Again, models satisfying the primary conditions
can be build, even a trivial one loop model may do.
In the trivial models, however, the mixing between the tau neutrino
and the sterile state would be large ($\sin^2 2\theta_\tau \sim 0.2$),
naively conflicting with the limits from the big bang nucleosynthesis.

\item {\em Karmen anomalies.}
The preliminary results of the Karmen experiment display an unexpected time
structure of the events \cite{Armbrusteretal95}.
These observations have already been interpreted as a
signal for a sterile neutrino of 33.9 MeV, with a mixing
to muon and electron neutrinos \cite{BargerPhillipsSarkar95}.
Below I will discuss some scenarios for such a particle, and present a more
consistent choice of parameters.

\item {\em Crisis in Big Bang Nucleosynthesis.}
The estimations of the abundances of the light elements in the universe
constrain the effective number of the degrees of freedom during the
nucleosynthesis. When expressed as a fit to the
number of neutrino species the results suggest there being
less than three neutrino species \cite{OliveSteigman95,Hataetal95},
contrary to
the LEP results [Eq.~(\ref{LEP})] that
state firmly the existence of three neutrino families.
Despite possible systematic uncertainties in both the
observations and the modelling, the results are quite
worrying.
Below I will discuss the possibility
to improve the fit with sterile neutrinos.

\item {\em Origin of baryonic asymmetry.}
The physics of neutrinos is closely related (see e.g.
\cite{PeltoniemiValle93a,ClineKainulainenOlive93a,Vilja94,%
Amelino-CameliaPisantiRosa94} and
references therein)
to the generation of the baryonic asymmetry of the universe. Usually neutrinos
do
not, however, any good for it, the
interesting possibility that the oscillation
to sterile neutrinos would be responsible for creating a lepton asymmetry
\cite{KhlopovPetcov81} is shown to work in the opposite direction, washing out
any deviation from the symmetric situation
\cite{EnqvistKainulainenMaalampi90}.
Also, requiring the baryonic asymmetry to survive the electroweak phase
transition sets stringent constraints for models of neutrino masses, e.g.
forbidding any Majorana masses above 10 eV \cite{ClineKainulainenOlive93b}.
Since the origin of the baryonic asymmetry is still poorly understood it
is premature to draw any definitive conclusions for the properties of
neutrinos.

\end{enumerate}
Beside these there have been several irregular results from different
beta decay experiments. At present the irregularities are not strong
enough to point to new physics.  Also many of the above anomalies
may be due to erroneous interpretation of the measured data or
defects in the  theory describing the phenomenon.

\subsection{Reconciling Los Alamos neutrinos with solar and atmospheric
neutrinos}
\label{recon}

With the three left-handed neutrinos one can find a mass matrix
solving simultaneously both the solar and the atmospheric
neutrino problems. Typically that kind of solutions yield neutrino masses
too low to have any cosmological significance. To generate masses
in the dark matter scale, while insisting on both the solutions, one
necessarily needs fine tuning which can be intelligently incorporated
within a specific model.
The new mass scale given by the Los Alamos experiment is completely
incompatible with solar neutrinos, unless neglecting at least one experiment
\cite{Kangetal95}.
The scale is also disfavored by the atmospheric neutrino solutions.
However, neglecting the results based on
stopping/passing ratio, there could be
a marginal possibility of having a muon neutrino
of about 0.7 eV, the other two neutrinos being very light,
 so that the solar neutrino problem would be solved by
the conversion from electron neutrino to tau neutrino,
the atmospheric neutrino
problem by oscillation from muon neutrino to tau neutrino, and the
oscillation from muon neutrino to electron
neutrino is then responsible for the
LSND
events. The required mass matrix would be very complicated, and intelligence
is required generating it without fine tuning.

For any other mass hierarchy
there is no way to fit with three neutrinos the
Los Alamos mass scale without abandoning either solar
or atmospheric neutrinos.

Sterile neutrinos have been suggested long ago as a solution
to this schism. Before the LSND results there were around three
possibilities to yield $4\times 4$ mass matrices consistent with
solar neutrino, atmospheric neutrino and hot dark matter requirements:
\begin{enumerate}
\item Muon and tau neutrino form a degenerated quasi-Dirac
neutrino with mass in the dark matter scale. The atmospheric
neutrino problem is solved by the oscillation from muon neutrino
to tau neutrino. Electron neutrino is light, with mass less than
$10^{-4}$ eV, and the solar neutrino  problem is solved by electron
neutrino converting to a sterile state with mass about $10^{-3}$ eV.
\item The solar neutrino problem is solved by oscillation from
electron neutrinos to tau neutrinos which are light. The atmospheric
neutrino deficit is due to muon neutrinos transforming to sterile neutrinos.
These neutrinos form a quasi-Dirac neutrino in the dark matter scale.
\item The electron, muon and tau neutrinos are light, the mixings
between them solve both the solar and the atmospheric neutrino
deficits. The sterile neutrino solely provides the hot dark matter.
\end{enumerate}
Of these, the second possibility is in imminent contradiction with
the oscillation limits from the early universe, while the third
one is strongly constrained by them.

Of the above three scenarios the first and the second are
compatible with the new mass scale suggested by Los Alamos results,
the third being disfavored.

The above arguments do not yet require more than one sterile state.
However, depending on the model it may be more natural to have
three. Moreover, more states may give more freedom to the theory,
and, above all, they may help to avoid some constraints that have
been put over the combined solutions.

With a second sterile state one gains another option for the
transition schemes: The solar neutrino problem is
solved by $\nu_e \to \nu_S$, these being light. The atmospheric
neutrino problem is then due to $\nu_\mu \to \nu_S'$, these
particles being quite degenerated, in the hot dark matter or LSND mass scale.
The tau neutrino, and a possible
third sterile state are left free, and can be used to avoid some
conflicts with other phenomena.

\section{Neutrino mass matrices with sterile neutrinos}
\label{models}

\subsection{Generalities}
The approach to study light sterile neutrinos disagrees with
the canonical view of the singlet states being very heavy.
Although the masses of all neutral singlet states are arbitrary and
independent in the
 SU(2)$\times$U(1) gauge model, it is more natural to
 expect them to be of some characteristic scale of the
 underlying theory. The lightness of the ordinary neutrinos,
however, gives already evidence that very disparate mass
scales can coexist, and we can expect the sterile
neutrinos to be light for similar reasons as the ordinary
neutrinos.

Apart from inconceivably small Yukawa couplings, there are three possibilities
to generate naturally small neutrino masses, all thoroughly
explored for the active sector: the see-saw mechanism, radiative
mechanisms and a low scale vacuum expectation value for the
higgs boson generating the masses. (See \cite{GelminiRoulet95} for a recent
review.)
All these can be applied almost identically to the sterile sector, but
we also want to generate a mixing between sterile and active neutrinos.

It is also possible that the final neutrino mass matrix is
due to several different mechanisms. Such a {\em hybrid
model} may include more degrees of freedom, allowing one
to generate more or less naturally any kind of mass
scenarios. Due to a large variety of possible neutrino
mass matrices, in the following sections we take the neutrino
masses and mixings as free independent parameters,
with no region being theoretically excluded.

The number of sterile neutrino species is generally free. The most obvious
choice is to assume that there is one per family, but some scenarios
may require them two. It is also possible that sterile neutrinos appear
outside the family structure, especially if they are originally
non-neutrinos, in which case their number is more arbitrary.

Throughout this work, I denote the standard left-handed neutrino weak
eigenstates by
$\nu_L$; the right-handed states, as appearing in Dirac case, or in
see-saw scenarios by $\nu_R$; and any exotic neutral singlet fermion
as $\nu_S$. The mass eigenstates, with the dominant component being
doublet or singlet state, respectively, are denoted by $\nu_A$ and $\nu_P$.
In some occasions the mass eigenstates are labeled by numbers, not referring
to their contents of weak eigenstates, so that $m_i < m_j$, if $i < j$.
I use $\theta$ for the mixing between  sterile and active states and
$\phi$ for the mixing between active neutrinos. The angle $\eta$ is
reserved for scalar boson mixing.

To protect the sterile neutrinos from having large masses,
it is assumed that there is a new symmetry,
denoted by S.
Since there is a plethora of different possibilities to choose
the symmetry group S, most of them behaving sufficiently similarly
in the considered phenomena,
I do not restrict to any particular symmetry group.
Neither is fixed whether the symmetry S is global or local.
The most obvious possibility of S being due to a mirror symmetry
\cite{LeeYang56,MaalampiRoos90}
has been recently discussed in \cite{FootVolkas95,BerezhianiMohapatra95}.
The known
particles are
not required to carry any S-charge, but depending on the model they may do,
possibly in a non-trivial way.
The sterile neutrinos can then be given masses by
breaking the S-symmetry spontaneously.
In all cases the particles under consideration may not be exhaustive,
the model may predict a zoo of other particles, too.
Especially, in case of a local symmetry the cancellation of possible
anomalies may require the existence of other particles.

It is most natural to expect the neutrinos to be Majorana particles.
A Dirac kind of mass gives more freedom by not being constrained
by neutrinoless double beta decay searches. However, it does not provide
{\em per se} a new solution to any of the
previously discussed anomalies, since
the degeneracy of the masses prevents the oscillation to sterile states.
One can break the degeneracy
by inducing a splitting between the mass eigenstates
by a small controllable Majorana mass term for
either the active or the sterile sector.  Since the protecting symmetries
should be already broken such terms are protected only by the
non-existence of scalars in appropriate representations.
A small splitting leads to a quasi-Dirac case, favorable for certain solutions,
especially to the atmospheric neutrino problem.  In some cases the quasi-Dirac
state may be also made of active neutrinos.

\subsection{See-saw scenario}

Taken that the see-saw mechanism is so beautiful for producing small neutrino
masses it is most natural to keep it, and expand it to the sterile sector.
Ignore the family mixing for a while, and assume that we have three
neutrinos, $\nu_L$, $\nu_R$ and
$\nu_S$.
When the symmetries are unbroken, only $\nu_R$ can have a
mass term (of Majorana type). Assume then that the S symmetry
is broken spontaneously, similarly with the symmetry breaking of
$SU(2)\times U(1)$. As a result of this there can arise a
mass term connecting $\nu_R$ to $\nu_L$ and $\nu_S$, so that
the tree level mass matrix looks like
\begin{equation}\label{seesawmass}
M = \left[\begin{array}{ccc}
0   & 0 & a \\
0   & 0 & b \\
a   & b & C
\end{array}\right]   ,
\end{equation}
in the $\nu_L, \nu_S, \nu_R$ basis. Typically it is assumed $a,b \ll C$.
Similar models, from a different point of view, have been considered recently
in \cite{Pilaftsis93,BamertBurgessMohapatra95},
it has also been shown that this kind of mass matrices are possible
in some supersymmetric scenarios \cite{ChunJoshipuraSmirnov95}.

Since the mass matrix (\ref{seesawmass}) is singular, there is one massless
state, defined as
\begin{equation}
|\nu_1 \rangle \simeq \sin \theta |\nu_L \rangle + \cos \theta |\nu_S\rangle.
\end{equation}
Of the two massive eigenstates the combination
\begin{equation}
|\nu_2\rangle \simeq \cos \theta |\nu_L\rangle - \sin \theta |\nu_S\rangle +
\epsilon |\nu_R \rangle
\end{equation}
is light, with a mass
\begin{equation}
m_2 = \frac{a^2+b^2}{C} = \epsilon^2 m_3,
\end{equation}
while the state
\begin{equation}
|\nu_3 \rangle \simeq |\nu_R \rangle
\end{equation}
has a mass $m_3 = C$.
The neutrino mixings above are defined as
\begin{eqnarray}
\tan \theta &=& \frac{b}{a}, \\
\epsilon &=& \frac{\sqrt{a^2+b^2}}{C},
\end{eqnarray}
where $\epsilon$ is supposed to be very small.

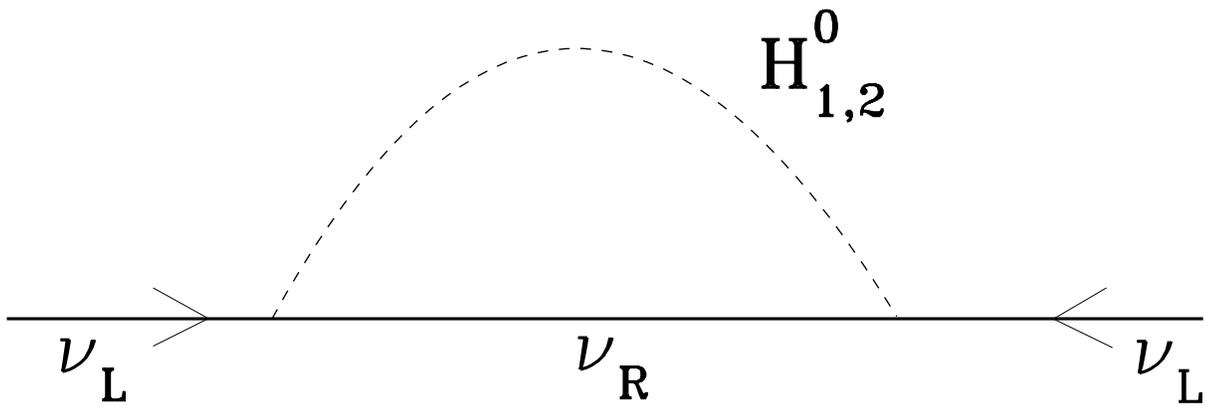
\begin{figure}
\caption{ The radiative corrections to the neutrino masses in the
extended see-saw scenario.
}
\label{kuvaMcorr}
\end{figure}

When higher order corrections are taken into account the
massless state may not remain massless any more.
As long as the mass terms $a$ and $b$ are generated by different non-identical
higgses there is no symmetry that can protect the masslessness.
Hence the graphs \ref{kuvaMcorr} give
\begin{eqnarray}\label{mac}
\delta m_{LL}
 &=& \frac{m_2 m_3^2}{32\pi^2 v^2} \cos^2 \theta
 \nonumber \\ &&
\left[
\left( \frac{M_1^2}{m_3^2-M_1^2}\ln\frac{m_3^2}{M_1^2}
- \frac{M_Z^2}{m_3^2-M_Z^2}\ln\frac{m_3^2}{M_Z^2}
\right)
\frac{ }{ }\right.
 \nonumber \\ &&\mbox{} \left.
 - \sin^2 \eta
\left( \frac{M_1^2}{m_3^2-M_1^2}\ln\frac{m_3^2}{M_1^2}
-\frac{M_2^2}{m_3^2-M_2^2}\ln\frac{m_3^2}{M_2^2}
\right)
\right]
 \nonumber \\ &&
+ \frac{g^2 m_2}{256\pi^2\cos^2\theta_W}
 \frac{m_3^2}{m_3^2-M_Z^2}\ln\frac{m_3^2}{M_Z^2}.
\end{eqnarray}
and
\begin{eqnarray}
\delta m_{LS}&=&
\frac{m_2 m_3^2 }{64 vw\pi^2}\sin 2\eta \sin 2 \theta
\nonumber \\ &&
\left( \frac{M_1^2}{m_3^2-M_1^2}\ln\frac{m_3^2}{M_1^2}-
 \frac{M_2^2}{m_3^2-M_2^2}\ln\frac{m_3^2}{M_2^2} \right)
\end{eqnarray}
where $\eta$ is the mixing between the higgs particles, and
$v$ and $w$ are the vacuum expectation values of the higgs
fields, and $M_1$ and $M_2$ are higgs masses.
The respective $\delta m_{SS}$ entry
depends on the details
of the symmetry S, it can be larger or smaller than the
other entries.

Typically the electroweak corrections induce to the would-be massless state
a mass
\begin{equation}
m_1 \sim \left( 10^{-3} \ldots 10^{-2}\right) m_2 \sin^2 2 \theta,
\end{equation}
while the loops involving the sterile neutrinos may give smaller
or larger contributions, depending on the model.
Only if there is a strict symmetry between sterile and ordinary neutrinos do
the different mass contributions cancel, conserving the masslessness.

In the three family basis we have still the freedom to choose
the number of the sterile states. The most natural choice
might be to assume that the above structure is replicated for
all families. Then we would have totally six singlet neutrinos,
and the mass spectrum could consist of three heavy states,
three intermediate states and three very light
states.
The families may mix as they do in normal see-saw scenarios,
with some more free parameters now. Non-trivial mass hierarchies
are also possible, e.g. one can have the mass eigenstates made
mostly of muon and tau neutrinos to be lighter than that dominated
by electron neutrinos.
The mixings
between active and sterile sectors are independent parameters,
not tightly connected to the physical masses.
However, it would be somewhat
unnatural to expect mixing angles very close to $\pi/4$, unless
there is some specific symmetry between standard and sterile sectors.
If the number of sterile states is different, the mass spectrum
will also change. For example, for only one $\nu_S$, there emerges
one almost massless state, and three light massive states.

\subsection{Radiative mechanisms}

With radiative mechanisms one does not need
heavy fermions
but instead new scalar particles are needed. The scalars, as well as the
fermions, need to have well defined
symmetry properties.  In several ad hoc models,
especially if requiring minimal
particle contents, the symmetry assignments tend to be non-trivial.

The simplest mechanism to generate radiative masses to standard
neutrinos is provided by the Zee model
\cite{Zee80,Zee85,Petcov82,Wolfenstein80b,SmirnovTao94}.
With one new singlet and another doublet, the one loop terms generate
a mass matrix leading typically to one heavy quasi-Dirac state, and
one light Majorana state. Hence this model gives naturally a mass spectrum
favored by the solutions to the atmospheric neutrino deficit while keeping
the lightest state (typically dominated by electron neutrino) sufficiently
light, just waiting for a sterile neutrino
to solve the solar neutrino problem.
For natural values of the parameters one cannot
obtain for ordinary neutrinos masses higher than 100 keV
without introducing new
fermions.

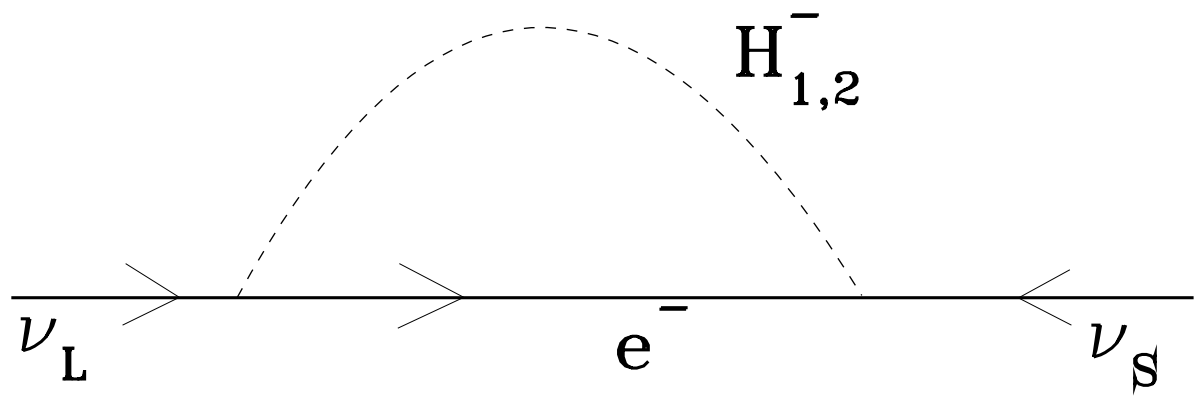
\begin{figure}
\caption{The graphs generating a mixing between doublet and singlet neutrinos.
}
\label{kuvaD}
\end{figure}

The mixing between ordinary and sterile neutrinos
can be generated by the one loop graph
presented in figure \ref{kuvaD}, with an ordinary charged lepton inside the
loop. Here
two new singly charged singlet scalars were introduced, one of which  carries
the same S quantum
number as the singlet state, and the other may be the singlet appearing in the
Zee model.
The scalars mix, due to the
spontaneous breaking of the symmetry, producing two mass eigenstates.
The mixing term
\begin{equation}
m = \frac{fh m_\ell }{32 \pi^2} \sin 2\eta
\ln\frac{M_1^2}{M_2^2},
\end{equation}
where $f$ and $h$ are Yukawa coupling constants, can reach values up to MeV
scale.

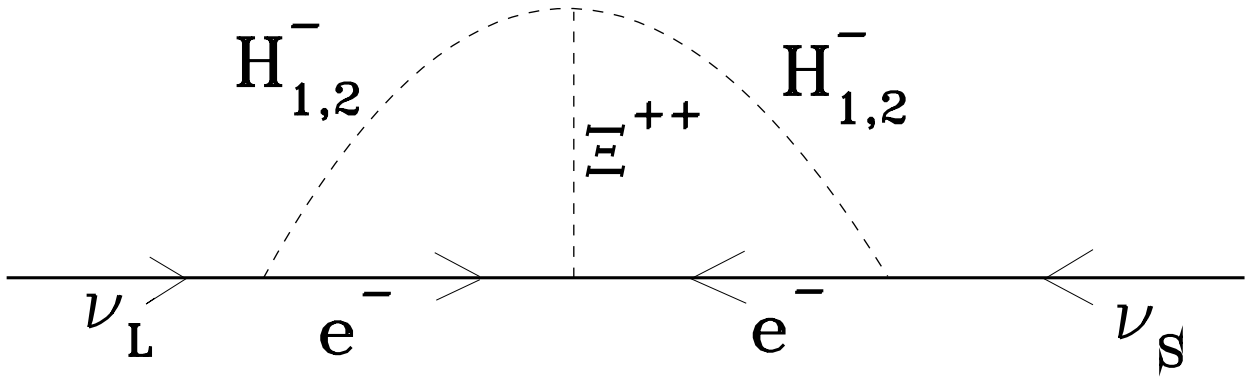
\begin{figure}
\caption{The graph generating a Majorana mass term for singlet neutrinos.
}
\label{kuvaDC}
\end{figure}

A Majorana mass term to the singlet states can be generated by the two loop
graph
\ref{kuvaDC}, with a new doubly charged singlet. The respective mass
element is given by
\begin{equation}
m_{SS} = \frac{f \zeta f D}{1024 \pi^4} \sin^2 2 \eta I\{ M\},
\end{equation}
where $f$ and $\zeta$ are Yukawa couplings,
$D$ is a scalar coupling constant of
dimension of mass, and $I\{M\}$ is a Feynman integral.
For typical values of the
parameters one can reach masses of O(keV) if insisting staying in the
electroweak mass scale. With other assumptions,
or different particle contents,
much larger masses are possible. One can also generate the sterile neutrino
mass terms
at one loop level, by introducing other particles
with more complicated quantum
number assignments.

With a more elaborated symmetry and particle contents one can obtain
more specific mass matrices. Such scenarios have been considered
abundantly in the literature, typically connected to multianomaly solutions.

One example is the model considered in
\cite{SmirnovValle91,PeltoniemiSmirnovValle92,PeltoniemiValle93b}.
It was based on the symmetry
$L_e+L_\mu-L_\tau$ (or $L_e - L_\mu + L_\tau$),
with the sterile neutrino having
lepton number $-1/2$. Several scalar particles were introduced. The resulting
neutrino mass
spectrum consists of a quasi-Dirac state of mainly  muon and
tau neutrinos, the other two mass eigenstates being much lighter.
The purpose of the model was to
reconcile the hot dark matter (initially the 17
keV neutrinos) with
solar and atmospheric neutrinos. Written before the LSND experiment,
it fits their observations perfectly.
Recently similar scenarios with different symmetries have been considered also
by \cite{CaldwellMohapatra93,MaRoy95}.

\subsection{Tree level masses}

A low vacuum expectation value of the higgs field generating the
neutrino masses can allow the masses to arise at tree level.
This way one avoids the need for additional fermions, or
peculiarly charged scalars.
The low scale of the symmetry breaking includes
a theoretical caveat, by introducing another hierarchy problem.
There are, however,  ways to generate
naturally low vacuum expectation values at a second order process.

The tree level Majorana masses for ordinary neutrinos have been
considered in triplet higgs scenarios,
appearing typically in left-right models.
To generate a mixing to sterile neutrinos one needs a doublet higgs,
necessarily different from the standard higgs generating the charged lepton
masses.
One can also have a singlet higgs inducing a Majorana mass term for singlet
neutrinos. In that case there is no need for a mass term for the left-handed
neutrinos, or the triplet field.

Since the symmetry breaking scale is typically connected to the mass
of the higgs bosons, this kind of models often contain light scalars.
That is not necessarily contradictory to any experiments, if the new
scalars couple only to the sterile sector. Assuming the respective neutrino
mass scales to extend below 1 eV, the natural mass scale for the scalars
lies below MeV. However, to generate a 30 MeV neutrino the respective
vacuum expectation value of the higgs field hardly can be lower.

\subsection{Large mixing and large hierarchy}

Some observations may hint for the existence of neutrinos at MeV scale.
Moreover, in the scenarios explaining those observations it is profitable to
have
a large mixing between the heavy state, and another much lighter state.
However, trivial mass matrices fail to give it, without fine tuning one
cannot obtain mixings much larger than $\sin^2 2\theta \sim 10^{-6}$
between neutrinos of 10 eV and 10 MeV.

Generally, models with specific symmetries can produce even massless states
with a large mixing to massive states.
One example is the above mentioned $L_e - L_\mu + L_\tau$ symmetry,
which can be applied also to sterile sector, like
$L_L - L_S + L_R$. That would lead to a mass matrix of the form
\begin{equation}\label{lsymmass}
\left[
\begin{array}{ccc}
0 & 0 & A \\
0 & 0 & B \\
A & B & 0
\end{array}
\right].
\end{equation}
This results in one massive Dirac neutrino state, and one massless Weyl state.
As long as the given symmetry is strictly in force,
radiative corrections cannot
induce any mass to the massless state.
A light mass term, as well as a splitting
between the Dirac partners can be allowed by a small breaking of the symmetry.
The considered mass matrix has the additional advantage that it is not
restricted
by neutrinoless double beta decay searches.
Such mass matrices, with the massive component very heavy have been considered
in \cite{WylerWolfenstein83}, but here we want also the heavy state
to be light enough.  Hence it is feasible to expect the masses to be due to
radiative
mechanisms.

Another form of a mass matrix leading to massless states mixing with massive
states is
\begin{equation}
\left[
\begin{array}{ccc}
aa & ab & ac \\
ba & bb & bc \\
ca & cb & cc
\end{array}
\right],
\end{equation}
which can be generalized to $n$ dimensions.  This results in one massive
Majorana state, and $n-1$ massless states. In some radiative scenarios
such mass matrices can emerge naturally, at least approximately.
Also here one requires a strict symmetry to keep the massless states really
massless. It is difficult, however,
to construct such a symmetry between doublet
and singlet neutrino states, and it may not necessarily commute with the
standard model
symmetry group. A special case of the above mass matrix, with $a=b=c$ is
often called democratic (or communistic), because of an obvious analogy:
in the initial basis all the components are equal, but in the physical
eigenstate basis one component is everything and the rest are nothing.

\section{Phenomenology of sterile neutrinos}  \label{pheno}

\subsection{New interactions}

The sterile neutrinos are assumed to have
interactions among them, as determined by the S-symmetry. For global and local
symmetries the
interactions
can be mediated by scalar bosons.
If S is a local symmetry then there are also gauge interactions, mediated by
new gauge bosons.
The effective scale of the interactions is not fixed, it can be essentially
higher or lower than that of the standard weak interactions, even though
it is most natural to assume the same scale. Hence the interactions
between the sterile neutrinos can be of the same magnitude or much stronger
than the standard
weak interactions.

If the interactions are mediated by a (gauge) boson with mass $M$,
the effective interaction strength is typically
\begin{equation}
G_S \simeq \frac{g_s^2}{q^2-M^2},
\end{equation}
where $q$ is the momentum change, and
$g_s$ is the coupling constant of the boson to the sterile neutrinos (S-charge
of the sterile neutrinos).
For scalar mediated interactions
it should be replaced by the respective Yukawa
coupling.
For low values of $M$, the interactions at high energies are suppressed as
$E^{-2}$. This can weaken the laboratory bounds discussed below. For
a specific case $M\sim 100$ MeV, a resonance may occur at energies
typical for neutrinos in the supernova core. A particle with such a mass
is not necessarily contradictory to anything, if its couples mainly to the
sterile sector. It is also heavy enough not to be produced excessively in
many environments.

The (gauge) interactions
in the sterile sector can lead to scattering processes
like
$\nu_P \nu_A \to \nu_P \nu_A$, $\nu_P \nu_A \to \nu_A \nu_A$,
or $\nu_P \nu_A \to \nu_P \nu_P$.
Such processes could also be mediated directly by scalar interactions,
very likely to be present in several models.
In model with radiative generation of masses, also interactions
$\nu_P \ell^- \to \nu_A \ell^-$ or $\nu_P \ell^- \to \nu_P \ell^-$
are likely to occur.
In some scenarios also the processes
$\nu_P q \to \nu_P q$ are possible
\cite{BabuMohapatraRothstein92}.
The sterile neutrino interactions can be theoretically almost as strong as the
standard model charged current interactions.
The considered scatterings can lead
to the production (or destruction)
of sterile neutrinos in several astrophysical
environments,  they can even cause a thermal or chemical equilibrium
between sterile and active neutrinos.

\subsection{Laboratory constraints}

The laboratory bounds on the sterile neutrino mixing depend on the mass scale,
the
mixing
of very heavy sterile neutrinos has been recently considered in
\cite{NardiRouletTommasini94,NardiRouletTommasini95,%
BarenbaumBernabeuJarlskogTommasini95}.
For sterile neutrinos in the few MeV scale the most stringent limits come
from searches of distortions in weak decays (see \cite{PDG94} and references
therein).
For example,
for 10 MeV one has the constraint $\sin^2 2\theta_e < 4 \times 10^{-4}$, while
for 33 MeV
$\sin ^2 2\theta_\mu < 0.008$ and  $\sin^2 2\theta_e < 3 \times 10^{-6}$.

For relatively light neutrinos,
the most relevant limits are from disappearance
experiments. For electron neutrinos, with a sufficiently large mass difference,
typically the reactor experiments give upper limits $\sin^2 2\theta_e < 0.14$
\cite{Zacecketal86,Vidyakinetal90}. These limits are not valid for masses
higher
than O(MeV),
also the mixing of neutrinos with $\delta m^2 < 0.008 \mbox{ eV}^2$ is
unconstrained.
The BEBC accelerator experiment yields $\sin^2 2\theta_e < 0.07$
\cite{Erriquezetal81}, for sterile neutrinos heavier than about 10 eV.
The upper limit for the mixings to the muon neutrino is less stringent,
typically $\sin^2 2\theta_\mu < 0.1$ for $3 \mbox{ eV}^2 < \delta m^2 < 1000
\mbox{ eV}^2$.
On the other hand, the mixing between tau neutrinos and
sterile neutrinos is practically unconstrained, for masses no higher than the
upper limits for the tau neutrino mass.

For heavy masses a stringent limit
for the mixing to electron neutrinos comes from
neutrinoless double beta decay searches \cite{Moe94}, that constrain
the mixing of any massive Majorana neutrino to electron neutrino as
\begin{equation}
\sin^2 \theta < \left(\frac{1 \mbox{ eV}}{m_\nu}\right),
\end{equation}
assuming different contributions not to cancel.
For neutrinos heavier than 10 MeV the above limit is weakened by kinematical
reasons,  for a 30 MeV neutrino by no more than a factor of 2.

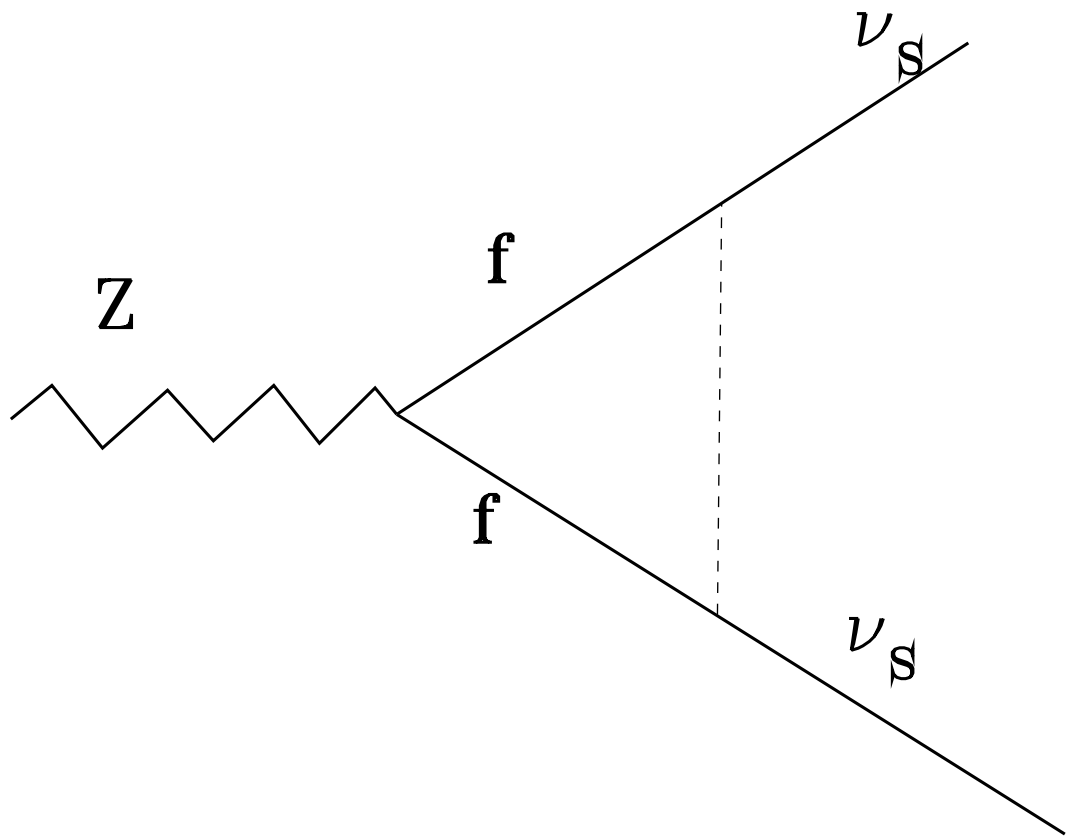
\begin{figure}
\caption{The one-loop graph inducing the Z decay to sterile neutrinos.
}
\label{kuvaZ}
\end{figure}

The laboratory experiments set also constraints for the new interactions
involving sterile neutrinos.
The constraint for the number of neutrinos from the Z decay width
\cite{PDG94}
\begin{equation}\label{LEP}
N_\nu = 2.983 \pm 0.025
\end{equation}
sets a bound for the effective four neutrino interactions
\cite{BilenkyBilenkySantamaria93}:
\begin{equation}  G_{\rm eff}(4\nu) < 4\times 10^2 G_F.
\end{equation}
This limit applies also to interactions between active and sterile neutrinos,
weakened by a factor of about 2.
More stringent limits can be derived by considering the one-loop
contribution to $Z \to \nu_S \nu_S$.
The graph depicted in Fig.~\ref{kuvaZ}, with an arbitrary fermion f in the
internal line, has a decay width
\begin{equation}
\Gamma_S = \left(\frac{g_f \xi_f \xi_f}{16\pi^2} \frac{M_Z^2}{M^2} I \right)^2
\Gamma_\nu,
\end{equation}
where $\Gamma_\nu$ is the decay rate to standard neutrinos, $I$ is a Feynman
integral, $\xi_f$ is the coupling between f and $\nu_S$, and $g_f$ is the
coupling of the fermion f to Z.
Assuming $I \sim $ O(1), we obtain the (approximate) limit
\begin{equation}
\frac{\xi_f \xi_f}{M} < \frac{200}{g_f} G_F.
\end{equation}
The limit does not apply for effective non-diagonal couplings $G_S(f f')$
(for which there are much stronger limits from rare decays),
neither for chirality flipping interactions of type
$G_S \bar{f}f\bar{\nu}_S \nu_S$.
Note that the limit given in
\cite{BilenkySantamaria94}
cannot be applied directly for sterile neutrinos.
The above limits can be avoided by fine tuning.
Since the neutrino mixing involves automatic fine tuning
one cannot draw conclusions on mixing angles.

\subsection{Neutrino decay}

A heavy neutrino, composed of singlet and doublet weak eigenstates, is
necessarily unstable. The decay can take place via
standard charged or neutral currents, and there is no
GIM mechanism to suppress the $\nu_A - \nu_P$ matrix element.
Decays $\nu_A \to \nu_A'$ remain suppressed though not forbidden.

The Z exchange interactions lead to the decay mode
$\nu_2 \to \nu_1 f \bar{f}$, where $f$ is any
kinematically allowed fermion coupling to Z.
The neutrino lifetime can be expressed as
\begin{eqnarray} \label{smlifetime}
\tau_\nu &=& \frac{1536 \pi^3 }{N_f m_\nu^5 G_F^2 \sin^2 2\theta}
\nonumber \\
&\simeq& \frac{1 \mbox{ s}}{\sin^2 2\theta}
\left(\frac{10 \mbox{ MeV}}{m_\nu}\right)^5,
\end{eqnarray}
where $N_f$ is the effective number of allowed final states, in
the latter formula electrons and three types of neutrinos were assumed as
possible
final states.
These lifetimes may allow the existence of heavy neutrinos, in the MeV scale,
since they decay away sufficiently fast. On the other hand, a light neutrino
in the dark matter range is essentially stable.

The decay to $\nu_2 \to e^+ e^- \nu_1$ can take place via
both charged and neutral currents. For the typical values of parameters
the neutral current reaction is the dominant one, its rate being determined
by the largest mixing while the charged current channel depends on the mixing
to electron neutrinos.
Hence the branching ratio of the electronic decay mode is
\begin{equation}
B_e = 0.15.
\end{equation}
For the most interesting range of parameters the decays to other
charged particles are all kinematically blocked.

Second order processes can induce a decay to photons. The standard electroweak
corrections
generate a transition magnetic moment
\begin{equation}
\mu_\nu = {3 \times 10^{-12} \mu_B}\; {\sin 2\theta}
\left(\frac{m_\nu}{10 \mbox{ MeV}}\right).
\end{equation}
Since the radiative decay rate depends similarly on the mixing and on the mass
as the three fermion decay via neutral currents, the decay rates are
proportional
to each others. The radiative decay is always weaker, with a branching ratio
\begin{equation}
B_\gamma = 0.008.
\end{equation}
This is sufficiently large to cause observable effects for a large region of
masses.

In models involving new charged scalars the transition moments may be
enhanced. This is especially case in models where the neutrino masses
are generated radiatively. Hence, depending on the details of the model,
one can obtain
\begin{equation}
\mu_\nu \sim 10^{-7} \mu_B \sin \theta
\left(\frac{m_2}{\mbox{ 10 MeV}}\right)
\left(\frac{M}{\mbox{ 50 GeV}}\right)^2,
\end{equation}
where $M$ is the scale of scalar masses.

If the symmetry S is global, its spontaneous breaking results in the existence
of a massless Goldstone boson, to be denoted by $J$. Since in case of sterile
neutrinos we necessarily
have neutrinos with different S charges, the coupling of the Goldstone boson
to neutrinos is not diagonal in the mass basis. Hence the decay
$\nu_2 \to \nu_1 J$ can take place unsuppressed, leading typically
to life times
\begin{equation}
\tau_\nu \sim \frac{0.1 \mbox{ s}}{\sin^2 2\theta}
\left(\frac{10 \mbox{ MeV}}{m_\nu}\right)^3
\left(\frac{w}{10^{10} \mbox{ GeV}}\right)^2,
\end{equation}
where $w$ is the vacuum expectation value related to the breaking of the
S-symmetry.
Unless the scale $w$ is very heavy, all relevant neutrinos are very
unstable. To make a 10 eV neutrino stable one requires
\begin{equation}
w > \sin 2\theta  \; 10^{10} \mbox{ GeV}.
\end{equation}
Note, however, that the decay $\nu_A \to \nu_A' J$ may be strongly suppressed.

Because of the neutrino mixing the sterile neutrino (gauge) interactions can
cause the
decay $\nu_2 \to \nu_1 \nu_P \nu_P$. The decay rate
can
be much larger than that via neutral currents, with the lifetime
given by
\begin{equation} \label{nslifetime}
\tau_\nu
\sim \frac{1 \mbox{ s}}{\sin^2 2\theta}
\left(\frac{G_F}{G_S}\right)^2
\left(\frac{10 \mbox{ MeV}}{m_\nu}\right)^5,
\end{equation}
depending on the details of the sterile sector.
Hence one obtains more
freedom for the scenarios requiring a rapid decay.
Moreover, it can reduce the  dangerous branching ratio to
photons, since the sterile interactions may not contribute
to the photon coupling.

\section{Sterile neutrinos in a supernova}
\label{sn}

\subsection{Supernova nucleosynthesis}

It has been claimed \cite{Qianetal93,QianFuller94} that supernovae would put
stringent constraints for
the mixing from tau
or muon
neutrinos to electron neutrinos.
The argument goes as following: Being emitted from their respective
neutrinospheres, muon neutrinos have on average higher energies than electron
neutrinos, electron antineutrinos having energies in between.
A conversion from a muon neutrino to an electron neutrino would result in
electron neutrinos having higher energies than electron antineutrinos,
even though the fluxes of each species would be close to equal.
Under such circumstances the layers containing heavy nuclei would
turn proton-rich due to electron neutrino radiation, which would prevent the
r-process.
If one assumes that the r-process is the origin of the nuclei heavier
than iron, such conversions should be disallowed.
The derived limits \cite{Qianetal93}
would restrict strongly the neutrino mixing
in the
dark matter mass range, and
especially exclude the attractive scenario of having
a massive muon neutrino as hot dark matter and explaining simultaneously the
LSND events.

Previously many authors have suggested that an inverted mass
spectrum ($m_{\nu_\mu} < m_{\nu_e}$) would solve the contradiction
\cite{FullerPrimackQian95,RaffeltSilk95,CaldwellMohapatra95}.
In this case
there is no conversion
to electron neutrinos, instead muon antineutrinos would convert to electron
antineutrinos. It is disputable whether this is in agreement with the
observations from SN1987A. The mass matrices implementing this scenario would
be of quite peculiar form. Although there is
nothing wrong in such scenarios, models leading to such a mass spectrum tend
to be less attractive.

Here I focus on an alternative remedy using sterile neutrinos.
Let us assume that we have a muon (or tau) neutrino with a mass
O(eV), and two other much lighter neutrinos, one of which electron type  and
the other sterile.
The matter induced potentials for the electron neutrino and the muon
neutrino, respectively, are given by
\begin{eqnarray}\label{nuepot}
V(\nu_e) &=& V_0 (3 Y_e - 1 + 4Y_{\nu_e} ), \\
V(\nu_\mu) &=& V_0 (Y_e-1 + 2 Y_{\nu_e}),
\end{eqnarray}
where
\begin{equation}
V_0 = 18 {\rm \; eV} \left( \frac{\rho}{5\cdot 10^{17} \rm \; kg/m^3}
\right),
\end{equation}
and $Y$ are the net abundances, relative to the nucleon density. The neutrino
contribution is generally
negligible outside the core. In the free streaming region it is further
suppressed because most neutrinos fly to the same direction.
It was assumed that no lepton flavor violation has occurred.

A level crossing can occur for both $\nu_\mu \to \nu_e$ and
$\nu_\mu \to \nu_S$ if the
state consisting mainly of the muon neutrino is the heaviest
state.
For antineutrinos an inverted mass spectrum is required.
The potential difference for the transition from a muon neutrino to a sterile
neutrino is now
smaller than that for the transition from a muon neutrino to an electron
neutrinos, outside
the strongly neutronized inner part. Consequently, the level crossing
for the transition to sterile state occurs in an inner zone than that for the
transition to electron neutrinos.

As a result of this level crossing scheme, a muon neutrino escaping
from the protoneutron star may convert to a sterile state before reaching
the muon to electron neutrino resonance zone.
 With the data of ref.~\cite{Qianetal93} we
can estimate that the transition is adiabatic for the considered
range of parameters, if the mixing angle satisfies
\begin{equation}
\sin^2 2 \theta_\mu > 10^{-4} \ldots 10^{-3}.
\end{equation}
This condition is apparently in conflict with the cosmological bound,
but taken that a partial transition is sufficient they may be marginally
consistent.
The resonance zones are situated very closely, and if they
are too wide they overlap. One can estimate that
the width of the resonances is narrower than
the distance
between the resonance zones
if
\begin{equation}
\sin^2 2\beta < 0.02,
\end{equation}
where $\beta$ is the largest mixing angle.
The suggested mixing from the Los Alamos experiment is below this limit in the
dangerous mass range,
hence the overlap depends on the mixing to the sterile neutrino.
However, even when the sterile neutrino mixing would violate the above
bound, to make the resonances overlap, it is expected that the neutrinos
converted to electron neutrinos are always a minority, as long as the
mixing to sterile states is not essentially smaller than the mixing
between electron and muon neutrinos.

\subsection{Sterile neutrino production in the core}

For neutrino masses below 100 eV the conversion to sterile states
in the core of the supernova is blocked by the strong matter effect.
The above scenario thus does not conflict the present picture of the
dynamics of the supernova. For sterile neutrinos with higher masses,
relevant for the discussion in the cosmology section, one can rule
out a certain area in the mass-mixing plane.
So far only the transitions from electron neutrinos to sterile neutrinos
have been considered seriously in the literature
\cite{KainulainenMaalampiPeltoniemi91}.
For $\delta m^2 \gg 2\times 10^9
{\rm \; eV^2}$ one can exclude
\begin{equation}  \label{snmix}
10^{-8} < \sin^2 2\theta_e < 0.02 ,
\end{equation}
and for  $\delta m^2 \ll 2\times 10^9
{\rm \; eV^2}$ the excluded region is given by
\begin{equation}
3\times 10^{5} \mbox{ eV}^2< | \delta m^2 \sin 2\theta_e |
< 3 \times 10^{8} \mbox{ eV}^2.
\end{equation}
The limits are very model dependent, the above constraints have
been chosen to be conservative. Within specific models of the supernova
interior one can obtain
limits that are stronger by an order of magnitude.
If a sterile neutrino has a large mixing with one active
neutrino, above the excluded region, its mixings with other active neutrino
species are not constrained.

For other neutrino species the respective limits may be different.
If muon and tau neutrinos are excited only thermally, with a zero
chemical potential, their abundances are much lower than those of
electron neutrinos. The conversions to sterile neutrinos from non-degenerate
neutrinos are not
irrelevant, however, they may be equally or even more
dangerous. The reason is that the emission of the sterile neutrinos
emerging from thermal neutrinos rapidly cools the protoneutron star,
which, consequently, accelerates the emission of ordinary neutrinos.
Hence, even though the energy carried out by sterile neutrinos never
exceeds that carried by normal neutrinos, the cooling of the star is
enough to upset the dynamics.
The limits for the conversion to sterile states should be
necessarily obtained by a self-consistent numerical simulation.
So far this has been done for estimating the allowed Dirac mass of neutrinos
\cite{BurrowsGandhiTurner92}, and the resulting limits are of the same
order of magnitude compared as those for degenerate neutrinos.

Similarly one can constrain the exotic
interactions producing sterile neutrinos.
The limits depend on the target particle.
For inelastic scattering from nucleons
($\nu_e N \to \nu_S N$) one can set the constraint \cite{RaffeltSeckel88}
\begin{equation}
G_S(N) < 10^{-4} G_F.
\end{equation}
Since electrons are very degenerate, their interactions are suppressed by
Pauli's rule, because most of the possible final states are occupied.
 The blocking factor
 $B_F \simeq \exp (-(E_f-\mu_e)/T)$, with $E_f$ the final electron energy,
$\mu_e$ the electron chemical potential and $T$ the temperature of the core,
may vary between $10^{-3}$ and 0.1 for temperatures between 10 and 40 MeV.
 Hence the respective limit for
the scattering from electrons ($\nu_e e^- \to \nu_S e^-$) is weaker,
\begin{equation}
G_S(e) < 6\times 10^{-4} G_F \ldots 6 \times 10^{-3} G_F.
\end{equation}
This limit is also more uncertain since it depends (exponentially) on the
temperature of the core which is a dynamical variable.
Even more uncertain are the limits for neutrino-neutrino scatterings.
For a process ($\nu_e \nu_e \to \nu_e \nu_S$)
or ($\nu_e \nu_e \to \nu_S \nu_S$)
one may set
\begin{equation}
G_S (\nu_e) < 0.001 G_F \ldots 0.01 G_F,
\end{equation}
while for the process ($\nu_e \nu_\mu \to \nu_S \nu_\mu$), assuming
non-degenerate muon neutrinos, the upper limit is about
\begin{equation}
G_S (\nu_\mu) < 0.001 G_F \ldots 0.02 G_F
\end{equation}
I refrain from giving limits for the transformation of non-degenerate
neutrinos.

The same interactions, when sufficiently strong,
can cause the sterile neutrinos
to be trapped in the core. However, the sterile neutrinos can also be trapped
by elastic scattering processes like $\nu_S e^- \to \nu_S e^-$.
The respective production channel $e^- e^+ \to \nu_S \bar{\nu}_S$ is blocked
because of a very low abundance of positrons, unless the core temperatures are
very high, $T \gg 50 $ MeV, or the electron chemical potential is reduced due
to lepton number violation.
We can estimate that the interaction
is sufficiently strong to be in concordance
with the observations if
\begin{eqnarray}
G_S(N) &>& 0.1 G_F \\
G_S(e) &>& 0.6 G_F \ldots 6 G_F\\
G_S(\nu_e) &>& 1 G_F \ldots 10 G_F\\
G_S(\nu_\mu) &>& 2 G_F \ldots 20 G_F.
\end{eqnarray}
Also here the constraints for the lepton interactions involve more
uncertainties, depending very strongly and non-linearly on the conditions
in the core. The quoted limits are again conservative, to be safe values more
stringent by an order of magnitude would be desirable.

Assuming that the
interactions are mediated by a boson that is massless or sufficiently light
(less than about 200 MeV)
the above limits can be written as
\begin{eqnarray}
\xi_{\nu_S} \xi_N &>& 10^{-8}, \\
\xi_{\nu_S} \xi_{\nu_e} &>& 10^{-6},
\end{eqnarray}
where $\xi_f$ is now a diagonal Yukawa coupling constant
with the fermion f.
Also these limits involve uncertainties by an order of magnitude.

For very heavy neutrinos the transition is blocked kinematically.
Since the electron neutrinos have chemical potentials about 150--200 MeV
just after the bounce,
the bounds  for the conversion from degenerate neutrinos
are valid for neutrino masses up to 100 MeV, above it the limits are
weaker. Masses higher than 300 MeV are safe. On the other hand,
for the transitions
from thermal, nondegenerate neutrinos the kinematical barrier depends on the
temperature which may vary between 8 MeV and 80 MeV. A mass of few
tens of MeV may still be dangerous, while more than 100 MeV can be already
considered
safe. The conversions will stop when the temperatures drop low enough,
for the 33 MeV neutrino the temperature of the core would be driven to below 5
MeV, assuming free emission, which may hardly be an acceptable value.

Consider then the possible effects due to resonant transitions
between electron neutrinos and sterile neutrinos. The respective
potential (\ref{nuepot})
 may cross the zero several times in the vicinity
of the core, where the electron abundance may vary above and
below 1/3, during the deleptonisation process \cite{Peltoniemi91a}.
This can lead to successive conversions and reconversions
between electron neutrinos and sterile neutrinos.
The adiabaticity condition depends strongly on the model of
the supernova evolution, and can be also substantially affected
by the conversions themselves. Typically conversions can
occur for masses above 10 eV, or even smaller if neutrinos have a large
magnetic moment.
These conversions may confuse the observable neutrino spectrum,
thus affecting the conclusions made from the observed neutrinos.
They may also change the
dynamics of the explosion, even reviving the explosion \cite{Peltoniemi91a}.
Now there appears also the possibility of a successive transition
$ \nu_e \to \nu_s \to \nu_\mu$, plausible for the eV scale masses, especially
as suggested by the LSND results. If all the conversions are
sufficiently effective, it may result to a beam of high energy
muon neutrinos.
Since the current experiments are quite insensitive
to muon neutrinos we cannot make any definitive conclusions from the
observations
of SN1987A for such transitions.

\subsection{Flavor conversions and internal deleptonisation}

For large neutrino masses it is possible that flavor changing oscillations
among active neutrinos
become effective in the core \cite{MaalampiPeltoniemi91a}
(See also \cite{Turner92,RaffeltSigl93}). Again, for masses
below 100 eV the
conversions are blocked, but above that the conversions may be rapid
enough to cause a chemical equilibrium between neutrino species, depending
on the mixing \cite{Peltoniemi92}. The transmutation time scale can be
estimated to be
\begin{equation}
t_r \sim \frac{10^{-8} \mbox{ s}}{\sin^2 2 \phi_m},
\end{equation}
where $\phi_m$ is the matter mixing angle.
Here the transition time is determined by the collisions of neutrinos
with the background particles via charged current. Such interactions
can involve as final states electrons, in which case the Pauli blocking
suppresses the cross section, or muons, provided the total energy exceeds
the muon mass threshold (105 MeV). The reactions leading to muons
have initially larger available phase space, until a degenerate
sea of muons is formed. Neutral currents,
being flavor blind, do not affect the
neutrino flavor conversion,
except when the effective neutrino mixing changes in
the
collision because of an energy change \cite{RaffeltSigl93}.

A conversion time of 0.1 s can be reached
for neutrino masses above 40 keV  if the
mixing
satisfies
\begin{equation}\label{emumix}
\sin^2 2\phi \sim 10^{-7},
\end{equation}
and for neutrinos lighter than 40 keV, respectively,
\begin{equation}
\delta m^2 \sin 2\phi \sim 10^6 \mbox{ eV}^2.
\end{equation}
Because of the non-linear dependence on energy and temperature, the above
results involve an uncertainty of an order of magnitude.
To reach a full equilibrium
 one may need much more time, since there may be other additional
transitions to the initial state or from the final state. In many occasions a
partial transition may be sufficient to cause significant effects.

As long as muon and tau neutrinos have equal distributions it makes no sense
to consider conversions between them. However, in case there is a flavor
conversion from electron neutrinos to one of them, they are no longer equal.
For the transitions between muon and
tau neutrinos there is initially no such a
large
potential difference to inhibit the transitions. Note that charged current
interactions can distinguish muon and tau neutrinos even if the matter
is symmetric between them, when the neutrino energies are above the muon
mass but below the tau mass.
Hence conversions among muon and tau neutrinos
with time scale of 0.1 s
 may be caused approximately by the mixing (\ref{emumix}).
However, for $\delta m^2 < 500 \mbox{ eV}^2$
the oscillation length exceeds the typical mean free path which damps the
oscillation. Hence, in that case,
to obtain the given transition time one needs
\begin{equation}
\sin^2 2\varphi (\delta m^2)^2 \sim 0.01 \mbox{ eV}^4,
\end{equation}
assuming no potential difference between muon and tau neutrinos. However,
if the conversion rate between muon and tau neutrinos is smaller than
the conversion from electron neutrinos to either of them,
it cannot maintain the
equilibrium and a
neutrino potential appears that can slow the conversion (or accelerate, if
there is a resonance) for small $\delta m^2$.
Also excited muons may contribute
to the effective potential. Moreover, several models generate a potential
difference between muon and tau neutrinos even in flavor neutral medium at
higher orders in perturbation theory \cite{Roulet95}.

A neutrino being in chemical equilibrium with the electron neutrino has the
same chemical potential
and is equally abundant as it.
The limits
for the mixing between this neutrino and a sterile neutrino are
then the same as those for electron neutrinos. Note that the
chemical equilibrium may change substantially the conditions in the core
\cite{MaalampiPeltoniemi91a},
as well as the dynamics of the shock wave, so that one
should treat the results with some caution.

Flavor equilibrium can be also induced by exotic flavor changing currents,
with a strength
\begin{eqnarray}G_{\rm FCNC}(N) &>& 10^{-4}G_F,
\\
G_{\rm FCNC}(\nu) &>& 10^{-2}G_F,
\end{eqnarray}
the first bound being valid for interactions with nucleons and the last
one for neutrino-neutrino interactions.

A large transition magnetic moment may cause an equilibrium
between electron neutrinos and antineutrinos of another flavor.
For this, one needs a magnetic moment of
$10^{-11} \mu_B$, obtainable in some radiative models.

A simultaneous occurrence of both flavor and spin flavor transitions
may result in a gross internal deleptonisation of the core.
The lepton number violation can also occur because of a Majorana
mass: the neutral current scattering can cause substantial helicity flip
(``neutrino-antineutrino transition'') for Majorana masses higher than 20 keV.
If any neutrino in flavor equilibrium with electron neutrinos has such a mass,
this will lead to the deleptonisation.
A majoron interchange process $\nu_e \nu_e
\to \bar{\nu}_e \bar{\nu}_e$ would have similar effects if the majoron
coupling satisfies $\xi > 5 \times 10^{-6}$.

The internal deleptonisation would invalidate all the limits previously
derived for neutrino conversions.
Under such
circumstances, the chemical potentials of all neutrino flavors would be driven
to zero. This would also accelerate the neutronization process $e^- p \to
n \nu_e$. As a consequence,
the temperature will rise rapidly, which
then changes the densities and the pressure.
The effects for the dynamics would be so drastic that it is impossible
to say anything quantitative without a new numerical simulation.

\subsection{Neutrino decay in supernovae}

The observed supernovae put restrictions on the decay of the
emitted neutrinos, and these apply to sterile neutrinos as well.
Especially the radiative decay of a heavy neutrino between the supernova
surface and the earth is strongly constrained. The non-observation of any
gamma-ray pulse \cite{SMM,KolbTurner89,Oberaueretal93} sets a bound
\begin{equation}
Y_R B_\gamma < 3\times 10^{-10},
\end{equation}
where $B_\gamma$ is the branching ratio to photons, and $Y_R$ is
the abundance of the heavy state in the surface of the star, relative to
stable standard neutrinos. For the standard electroweak decay modes
with sterile neutrinos, discussed in section \ref{pheno}, the branching
ratio to photons is never sufficiently low to satisfy the above limit, hence
their abundance must be  diluted kinematically, or they must decay inside
the star. Only if the neutrinos are
much heavier than 100 MeV is the kinematical
dilution sufficient alone. For lighter neutrinos a very small fraction of
neutrinos is allowed to survive up to the surface, so that for neutrino
masses about or less than 30 MeV the life time
should satisfy
\begin{equation}\label{snlifetime}
\tau_\nu < 10 \mbox{ s} \left(\frac{m_\nu}{30 \mbox{ MeV}}\right).
\end{equation}
The above limit is stronger, by about a factor of 2, if the neutrinos
interact more weakly, because of a smaller mixing for instance, so that
they are emitted from a hotter region closer to the center.
Note that there is no solution with (\ref{smlifetime}) for
$m_\nu < 8$ MeV.

The decay inside the star may deposit energy
to outer regions which may then have some consequences on the explosion
itself.
For the standard model decay channels
the energy transfer may be very effective
since 10 \% of the decay products involve electrons or positrons.
Taken that the observed kinetic energies are about 1 \% of the total
energy
released, a much larger energy transfer would be inconsistent unless the
absorbed energy is immediately transformed to new neutrinos.
In some scenarios it may be very profitable to bring new
energy to outer regions, helping
the star to explode. However, for the life times of 0.1 s -- 10 s the
energy may be transited too out (beyond 10 000 km from the center). To deliver
the energy on a region
of about 1000 km, a lifetime of O(5 ms) is required.
The total energy deposited to the envelope depends on the initial flux of
the sterile state which is typically from 0.1 to 10 times the flux of standard
neutrinos, depending on the interactions of the sterile state.

A decay to only neutrinos or other
neutral objects outside the core is harmless
for the dynamics.
A decay to freely escaping particles inside the supernova core, instead,
may be very dangerous, leading to a rapid energy flow out of the star.
Hence one may exclude (relativistic)
lifetimes less than about 0.1 s. Note that this limit is more conservative
than that given in \cite{NussinovMohapatra95}. For very short lifetimes
there are necessarily new interactions between active and passive neutrinos,
and these may be actually much more dangerous, though in extreme case
they may evade the bound by trapping.

\section{Cosmology of sterile neutrinos}
\label{bbn}

\subsection{Big bang nucleosynthesis}

A sterile neutrino with a substantial mixing to active neutrinos would
keep in thermal equilibrium with the active universe. This might
contradict the present results about the big bang nucleosynthesis and
the observed amounts of light elements.
The most recent limit for the effective number of neutrinos
$ N_\nu < 3.6$ \cite{OliveSteigman95}
forbids an additional light sterile neutrino, but
allows an additional light scalar boson. This bound
should be treated with
some caution, however, since actually the best fit for the effective number of
neutrinos lies deeply in the unphysical region:  Ref.~\cite{OliveSteigman95}
obtained
$N_\nu = 2.17 \pm 0.27 \pm 0.42$, while in \cite{Hataetal95} it was claimed
\begin{equation}\label{nenumb}
N_\nu = 2.0 \pm 0.3,
\end{equation}
excluding $N_\nu = 3$ at 99.7 \% confidence level. This hardly can
be a statistical fluke, but  may rather
hint that there is something that is not fully understood yet.
The poor consistency of the standard model detracts from the
credibility of the bounds for non-standard physics.

The constraints for the additional number of neutrinos have been
used to bound the mixing between active and sterile neutrinos
\cite{BarbieriDolgov90,Kainulainen90a,EnqvistKainulainenThomson92,%
ShiSchrammFields93,EnqvistMaalampiSemikoz95}.
These limits, taken at their face values rule out the sterile
neutrino interpretation
of the atmospheric neutrino deficit
with a clear margin. They do not yet contradict the sterile neutrino
solution to the solar neutrino problem, but
they would be detrimental to some scenarios discussed above.
Let us next study some scenarios to avoid such limits.

An apparently trivial way to avoid the oscillation to a sterile state is to
assume
that the mixing, or the splitting, arises only after the neutrino decoupling.
This could be arranged,
if the sterile neutrino masses and mixings, or the splitting, are due to a
spontaneous
breaking of a symmetry which occurs at temperatures below
O(1 MeV).
In the symmetric phase, at high temperatures, the neutrino
oscillation is then inactive, and the sterile states could be
excited only by interactions explicitly coupling to both
sterile and active sectors,
needed to generate the neutrino mixing. However, the
constraints for such interactions
are respectively very strong, and it is not self-clear that those can be
consistently satisfied without releasing the initial presumptions.
In fact, most scenarios fail to do this.
If the interactions occur via a heavy state, as in see-saw scenarios
the limits can be satisfied by construction.
Another possibility is that neutrinos are initially Dirac particles,
the splitting being induced in the late phase transition. In that case
all the new interactions concern only the sterile particles that are
sufficiently decoupled.

Another remedy is to induce a
large effective potential due to interactions mediated by majorons, or other
light particles \cite{BabuRothstein92}. For a certain range of parameters that
may
relax the constraints \cite{EnqvistKainulainenThomson92c}.
Also, a large $\nu_S$-$\bar{\nu}_S$ asymmetry might induce a substantial
potential term for the sterile neutrinos, blocking partially the transitions.

A more attractive escape may be provided by a heavy neutrino.
The density of a sufficiently heavy neutrino can be diluted kinematically,
by pair annihilation, so that it would not count as a full neutrino species.
For standard interactions one needs masses above 30 MeV. The recent
measurement for the tau neutrino mass ($m_{\nu_\tau} < 24$ MeV) \cite{Aleph95}
closes this window for tau
neutrinos. Nevertheless,
this scenario allows the existence of heavy sterile neutrinos, provided they
have sufficiently strong interactions. Note that this is a remedy for only the
nucleosynthesis limit, for other cosmological purposes the heavy neutrinos must
be essentially
unstable, unless they are heavier than several GeV.

Also a rapid decay may reduce the density sufficiently, and perhaps more
naturally \cite{Kawasakietal94}.
A decay to standard particles would not have any other side effects
for the nucleosynthesis, if it is sufficiently rapid. A decay to invisible
particles may help if the the heavy particle is already sufficiently diluted.

Other than only deleting one neutrino species, the neutrino decay can be
used to modify the onset of nucleosynthesis, to solve the conflict with the
result (\ref{nenumb}).
It has been shown \cite{GyukTurner94} that an unstable neutrino with mass of
the
order of 1 MeV, and a lifetime from 0.1 s to 10 s, decaying to electron
neutrinos (or light mixed neutrinos with a substantial component of
$\nu_e$) would balance the effect on the neutron to proton ratio.
The reason for this is that the resulting electron neutrinos,
with energies higher than thermal, would keep the neutron to proton
ratio longer in equilibrium, thus reducing the neutron fraction.
Since one then needs less neutron decay to fit the observed neutron fraction,
the expansion of the
universe in the considered times could be accelerated by additional
energy density. With certain values of the parameters as many as 16
new neutrino-like degrees of freedom would be allowed.
More than only allowing new objects, this mechanism would indeed give
a better fit to the observations, solving the apparent conflict between
big bang nucleosynthesis and standard particle physics model.

Previously mainly tau neutrinos have been considered as the heavy
state, since it is the only known neutrino allowed to have a sufficiently
heavy mass. As  the sterile neutrinos may liberate the tau neutrino from
the other duties, such a possibility remains very viable.
Furthermore, the sterile neutrinos open a valid channel for the
decay of tau neutrinos, namely $ \nu_\tau \to \nu_P \nu_e \bar{\nu}_e$,
via standard neutral currents,
with the life times (\ref{smlifetime}) being in the required region.
Unavoidably the decay products would include also other than electron
neutrinos since the neutral currents are flavor blind. There should,
however, be sufficiently many electron neutrinos to do the task.

Alternatively we can assume the heavy state to be mostly a sterile neutrino.
This choice actually gives us more liberty,
since the density of the sterile neutrino can be considered phenomenologically
as an
independent degree of freedom.
As above, the heavy state decays via neutral currents as
 $\nu_P \to \nu_A \nu_A \bar{\nu}_A$,
with the life time as above.
For $m = 10$ MeV and $\sin^2 2\theta = 0.02$ the life time
is about 10 s, just within the correct magnitude.
Defining the allowed parameter range would require a more
thorough analysis of the output on nucleosynthesis, most likely involving
numerical simulation, and it depends also on the full particle spectrum of the
model. This remains out of the scope of this work;
obviously there is sufficiently freedom to make an acceptable fit.

The relatively high mixings required for the fast decay may be consistent
with the limits from the supernovae, Eq.~(\ref{snmix}).
Nevertheless, it may be theoretically difficult to build a
natural scenario giving large mixings with large mass differences, and
an appropriate family hierarchy.
One remedy for the problem of a single heavy state with large mixing
is to assume the daughter neutrino to be also quite heavy. This is, of
course, only a partial solution since one has then to get rid of the
daughter neutrino also which is no easier, even though there are
less restrictions for the decay channel.

The decay may also take place via some new interactions.
A sufficiently strong interaction in the sterile sector may similarly
cause a rapid decay, the majority of the decay products being also
most likely sterile. A scalar interchange process can also lead to electrons.
In models
with a global symmetry the decay to
massless Goldstone bosons may be very rapid.
In typical models without coincidental suppression, to obtain a lifetime of 0.1
s for a 20 MeV neutrino
decaying to electron neutrino, the scale of the symmetry breaking
should be no more than $10^6$ GeV, not to violate the limits
from neutrinoless double beta decay.

\subsection{Dark matter and structure formation}

A lately decaying particle may pose some problems for the
initiation of the structure formation of the universe in some
models. However, in some other models, notably the
formerly popular cold dark matter scenario, this may be exactly
what is needed to make it work, since the decay may result in the increase in
the energy density of relativistic particles which, consequently, delays the
beginning of the matter dominated era
\cite{BardeenBondEfstathiou87,BondEfstathiou91,DodelsonGyukTurner94,%
WhiteGelminiSilk95}.
For a massive neutrino one needs a lifetime
 \begin{equation}
 \tau_\nu \simeq \frac{50 \mbox{ s}}{Y_\nu^2}
 \left(\frac{10 \mbox{ MeV}}{m_\nu}\right)^2,
\end{equation}
where $Y_\nu$ is the abundance of the heavy neutrino relative to
the standard neutrinos, after its decoupling.
For the mass ranges typically considered, from 100 eV to few MeV, the
standard model neutral current decay mode is insufficient. Hence one
has to introduce new interactions, a decay to a massless majoron is typically
appropriate. On the other hand, for several MeV masses the neutral currents do
well if the mixing is sufficiently large.

A sterile neutrino heavier than few electronvolts may provide
a substantial part of the mass of the universe. Since we assumed
the sterile neutrinos to mix with the ordinary neutrinos they
are likely to be excited during the evolution of the early
universe, so that their present number densities are not
negligible, unless they decay sufficiently fast to ordinary matter.
Whenever the mixing between  active and sterile neutrinos is
within the forbidden region of ref.~\cite{EnqvistKainulainenThomson92}, the
sterile neutrinos may be almost as abundant as the active ones.
Moreover, the interactions the sterile neutrinos under consideration
necessarily have with ordinary matter may also be responsible
for exciting them. These interactions may be specially
strong in models with radiative generation of neutrino masses.
In scenarios with heavy neutrinos, a decay $\nu_2 \to \nu_1 \nu_1 \bar{\nu}_1$
may increase the number of $\nu_1$ above the usual neutrino densities.
Hence, a sterile neutrino with a mass from 10 eV to 1 keV may
close the universe, heavier neutrinos must be unstable.

The effect of the relic sterile neutrinos on the formation of
the structures may be different from that of ordinary neutrinos.
Since the sterile neutrinos may be produced in a different way,
their temperatures may be also different, or they may even
have nonthermal distributions. The particles emerging from
the decay of a non-relativistic particle necessarily have
average energies higher than given by the thermal distribution.
Hence quite a heavy particle (a few hundred electronvolts) can behave as hot
dark
matter.
On the other hand, the particles decoupled very early are
colder than the rest of the universe, typically a particle decoupling
around the electroweak phase transition has a present day temperature
half of that of the visible universe.
As a curiosity, one
can build scenarios where the very same particle provides
both the hot and the cold dark matter, hence giving a more
literal sense to the expression.

Apart from some other dark matter candidates, the sterile neutrinos lack
the dissipative mechanisms needed to make them coalesce in galaxies.
Hence they are not optimal to form the galactic halos, even
if their masses would be large enough (more than 30 eV)  to avoid the
phase space constraints.

\section{Karmen anomalies} \label{karmen}

The preliminary results of the Karmen experiment \cite{Armbrusteretal95}
on the number of events as a function of the time after the pion collision
show an unexpected bump at
about 3.6 $\rm \mu s$. The results are not yet statistically
significant, to achieve that one needs more years of data
taking. However, it is interesting to speculate on the possible
particle physics origin of such a bump.

An evident cause would be a heavy particle produced via the two body
decay of the pion to a muon. That particle would then decay to something
visible
inside the detector.
The velocity of the hypothesized particle is determined to be
$v = 0.017^{ + 0.007}_{ - 0.005} c $,
and kinematically one can obtain for its mass a value of
33.9 MeV \cite{Armbrusteretal95}.
Since
the number of detectable events is a function of both the production
and the decay rates of the new particle,
one can make for $\tau > 10 \; \mu\mbox{s}$ the fit
\begin{equation}\label{btau}
B_P \sim \frac{3\times 10^{-11}}{B_{\rm vis}}
\left(\frac{\tau}{1 \rm \; s}\right),
\end{equation}
where $B_P$ is the branching ratio to the heavy particle in pion decay,
$\tau$ is the lifetime of that particle, and $B_{\rm vis}$ is the fraction of
the decays
of the new particle to something visible.
No known particle fits the measurements.

In \cite{BargerPhillipsSarkar95} it was suggested that the
culprit is a sterile neutrino, produced with the branching ratio
\begin{equation}
B_P = 7 \times 10^{-3} \sin^2 2\theta_\mu.
\end{equation}
Assuming it to decay
by charged currents
as $\nu_P \to e^- e^+ \nu_e$
they obtained for its mixings with the standard neutrinos the range
\begin{math}
10^{-9} < \sin^2 2\theta_e < 2.5\times 10^{-7}
\end{math}
and
\begin{math}
3\times 10^{-5} < \sin^2 2\theta_\mu < 8 \times 10^{-3}.
\end{math}
These values are inconsistent with the constraints from both the supernova and
the early universe.
The authors of \cite{BargerPhillipsSarkar95},
however, ignored the visible neutral current decay
modes.

Generally the dominant decay channel goes via neutral currents,
$\nu_P \to \nu_A f \bar{f}$, where $f$ can be a light neutrino or
an electron. The life time can be estimated
\begin{equation}
\tau_\nu \simeq \frac{2 \times 10^{-3} \mbox{ s}}{\sin^2 2 \theta},
\end{equation}
where $\theta$ is the largest mixing to any active neutrino.
(If the mixings are equal, then one should use the sum of the
respective squared mixing components instead).
The branching ratio to electrons is $B_{\rm vis} =0.15$.
Thus the equation (\ref{btau}) gives a relation
\begin{equation}\label{Karmensol}
\sin^2 2\theta_\mu \sin^2 2\theta \simeq 6 \times 10^{-11}.
\end{equation}
Assuming both the production and the decay to be determined
by the same mixing angle to muon neutrino, one obtains a unique solution
$\sin^2 2\theta_\mu = 8\times 10^{-6}$, and $\tau = 220$ s, which is
inconsistent
with supernova observations and dangerous for cosmology.
If the production and decay depend on a different mixings,
we have two free parameters, and a large range of lifetimes
is possible.
For instance, a life time of one seconds can be obtained
with the mixing angles $\sin^2 2\theta_\mu = 3\times 10^{-8}$
and $\sin^2 2\theta_\tau = 2\times 10^{-3}$.
Requiring a large mixing, e.g. $\sin^2 2\theta_\tau > 0.02$,
as suggested by (\ref{snmix}),
one obtains
\begin{equation} \tau_\nu < 0.1 \mbox{ s},
\end{equation}
and
\begin{equation}
6 \times 10^{-11} < \sin^2 2\theta_\mu
< 3 \times 10^{-9}.
\end{equation}
Additional invisible decay modes would not change the given solutions
(\ref{Karmensol}) for
the mixing angles.

The mass and the range of the lifetimes of the sterile neutrino suggested
by Karmen experiment are very interesting from a cosmological
point of view. Naively, the result (\ref{Karmensol}) is contradictory to the
cosmological limits for the mixing with sterile neutrinos. However, the
solution
itself provides the ingredients to avoid the given limits by a rapid decay. The
mass and
the lifetime of the heavy neutrino fall into the range where they naturally
provide the early decaying particle needed to solve the conflict with the
nucleosynthesis.

These results can also be in agreement with the
considerations of neutrino emission from supernova, including the limits
for the radiative decay. For the required mixing the sterile state would be
trapped in the core.
The neutrinos escaping the core, with the suggested properties,
may deliver dangerously much energy in the outer regions of the star, but for
sufficiently short lifetimes it
may be used for profit, to help blowing up the envelope. Note also that
the emitted flux of the heavy neutrinos may be reduced kinematically,
especially
if their interactions with matter are sufficiently strong (large mixing).
Any additional invisible decay mode would reduce the transferred energy.

Other solutions can be found with non-standard interactions. In such a case
the mixings can be minimal allowed by (\ref{Karmensol}),
$\sin^2 2\theta \sim \rm O(10^{-5})$.
The new interactions should then trap the neutrinos in the core of the
supernova,
and also cause the decay  of the heavy neutrino. To obtain lifetimes shorter
than 10 s, consistent with supernova and early universe considerations, at
least if the decay products are active neutrinos, one needs a nondiagonal
effective coupling $G_S > 0.01 G_F$.

\section{Conclusions}
\label{conc}

The sterile neutrinos can provide a solution to the solar neutrino
deficit, the atmospheric neutrino problem, the missing matter of
the universe, the anomalous ionization of interstellar hydrogen,
the explosion of a supernova, the conflict between a part of the parameter
space suggested by the LSND experiment and the supernova nucleosynthesis,
the crisis of the big bang nucleosynthesis, and
the anomalies observed in the Karmen
experiment. All these problems can also be solved individually
without sterile neutrinos, either by ordinary neutrinos or by some other
objects. It is also possible that some, if not all, anomalies are due to
misconception in underlying theory or uncontrolled phenomena in the
experimental
apparatus.

The standard electroweak model with three massive
neutrinos allows only a simultaneous solution of two or three of the above
problems, unless excessive fine tuning is applied.
By introducing three sterile neutrinos one can solve almost all of the
above problems simultaneously, as well as fit the LSND results.

The existence of light sterile neutrinos is not itself neither
theoretically inconsistent nor unnatural. Several effective models
can generate a small mass to sterile neutrinos, with a mixing
to active neutrinos. In this sense, the expression sterile neutrino
can refer to any exotic fermion without standard model interactions,
with a mixing to standard neutrinos.

The sterile neutrino solution to the atmospheric
neutrino problem is not contradictory
to any observation, and can be taken as a
viable alternative.
In many models it is more natural that the favored
quasi-Dirac state is made of a doublet and a singlet components than
that it were made of two active neutrinos. Hence, one can build a scenario
of using the sterile neutrinos to solve both solar and atmospheric neutrino
problems, and muon neutrino to electron neutrino oscillation to solve the
LSND results.

It was found that the conversion of a muon neutrino to a light sterile state in
the
supernova can solve the apparent conflict between the large mass solutions
of the LSND results and the supernova nucleosynthesis. This does not conflict
any bounds derived from SN1987A observations, or any other dynamical
consideration.
It was also recalled that the sterile neutrinos can help to blow up
the supernova, either due to a reconversion to active state, or
by decaying to electromagnetically interacting particles.

Limits for the production of sterile neutrinos in the supernova core
were reconsidered, and some new limits were presented.
These limits are modified in case of equilibrium
between neutrino flavors, or neutrino helicity states. The latter may result
in an internal deleptonisation of the core which upsets its dynamics
completely.
In such a case all the present considerations for neutrinos in supernovae would
cease to be valid.

The limits for light sterile neutrinos
in the early universe can be evaded by introducing more sterile neutrinos.
An early decaying heavy neutrino, with mass O(10 MeV),
may solve the nucleosynthesis anomaly, and may also
allow new additional light neutrino states.
To obtain the required lifetime
it is enough that the heavy mass eigenstate is composed significantly of
both active and sterile weak eigenstates.
It is emphasized that one cannot
rely very much on the constraints from the early universe for new physics
anyway,
since the observed abundances of the
light elements are inconsistent with the standard particle physics model
and the current models for the big bang nucleosynthesis.

The sterile neutrino solution to the Karmen anomalies fits the
observed results. It was found to be consistent with cosmology, and may
even improve the nucleosynthesis. It is also in
agreement with the observations of SN1987A, possibly helping the star to blow
up.
The presented parameter range differs from the
previous
speculations. Non-standard interactions may expand the allowed ranges for
the mixing angles.

The suggested solution for Karmen anomalies, together with other requirements,
may lead to an extraordinarily shaped mass matrix. The large mixing with
a heavy neutrino generally implies the other state also to be quite heavy.
A general $2\times 2$ mass matrix without fine tuning would imply the mass of
the lighter state in the
MeV range, assuming the heavy state to be 33 MeV,
while the see saw like mass matrix (\ref{seesawmass}) may predict masses
of the order of 17 keV. Only models involving a specific symmetry with active
and sterile neutrinos could produce mass matrices with a strong hierarchy
without fine
tuning. Alternatively, if the  decay occurs by explicit non-standard
interactions between
active and sterile neutrinos there is no need for large neutrino mixing.

The sterile neutrinos may provide part or all of the missing matter of the
universe. Due to the large variety of possibilities for their masses and
origins, they can act as either hot or cold dark matter, or even both.
The decaying neutrino scenarios can result in relatively massive neutrinos
being hot, while an early decoupled neutrino is colder. The relic density of
sterile neutrinos may be lower or higher than that of ordinary neutrinos. A
radiative decay of relic sterile neutrinos may be the
cause for the anomalous ionization of the interstellar hydrogen, which
does not require any new physics beyond what was considered here.


\setlength{\itemsep}{0mm}

\end{document}